\documentclass[acmtog,authorversion]{acmart}

\usepackage{booktabs} 

\usepackage[ruled]{algorithm2e} 

\SetAlFnt{\small}
\SetAlCapFnt{\small}
\SetAlCapNameFnt{\small}
\SetAlCapHSkip{0pt}
\IncMargin{-\parindent}

\acmJournal{TOG}
\acmVolume{36}
\acmNumber{4}
\acmArticle{45}
\acmYear{2017}
\acmMonth{7}

\acmDOI{http://dx.doi.org/10.1145/3072959.3073641}

\setcitestyle{square}

\usepackage{color}
\usepackage{graphicx}
\usepackage{epsfig}
\usepackage{amsmath}
\usepackage{amssymb}
\usepackage{multirow}
\usepackage{enumitem}
\usepackage{colortbl}
\usepackage{pgfplotstable}
\usepackage{ifthen}

\definecolor{dblue}{rgb}{0.0,0.0,0.5}
\definecolor{dgreen}{rgb}{0.0,0.5,0.0}
\definecolor{dred}{rgb}{0.5,0.0,0.0}

\newcommand{\ignorethis}[1]{}

\def\equationautorefname~#1\null{Equation~(#1)\null}

\renewcommand{\paragraph}[1]{\subsection{#1}}
\newcommand{\myitem}[1]{\item \textbf{#1}}

\def\roughness{{\alpha}}

\def\x{x}
\def\n{n}

\def\w{\omega}
\def\wi{{\w_i}}
\def\wo{{\w_o}}
\def\wh{{\w_h}}

\def\kd{{\rho_d}}
\def\ks{{\rho_s}}
\def\fr{{f_r}}

\def\E{E}

\renewcommand\dot[2]{{(#1 \cdot #2)}}

\def\pos{\x}

\renewcommand\exp[1]{{e^{#1}}}

\begin{document}

\title{Modeling Surface Appearance from a Single Photograph using Self-augmented Convolutional Neural Networks}

\author{Xiao Li}
\affiliation{\institution{University of Science and Technology of China \& Microsoft Research Asia}}

\author{Yue Dong}
\affiliation{\institution{Microsoft Research Asia}}

\author{Pieter Peers}
\affiliation{\institution{College of William \& Mary}}

\author{Xin Tong}
\affiliation{\institution{Microsoft Research Asia}}

\renewcommand{\shortauthors}{X. Li et al.}

\thanks{Part of this work was performed while Pieter Peers visited
  Microsoft Research Asia.\\}

\graphicspath{{figure/}}


\begin{abstract}
    We present a convolutional neural network (CNN) based solution for
    modeling physically plausible spatially varying surface
    reflectance functions (SVBRDF) from a single photograph of a
    planar material sample under unknown natural
    illumination. Gathering a sufficiently large set of labeled
    training pairs consisting of photographs of SVBRDF samples and
    corresponding reflectance parameters, is a difficult and arduous
    process.  To reduce the amount of required labeled training data,
    we propose to leverage the appearance information embedded in
    unlabeled images of spatially varying materials to self-augment
    the training process. Starting from an initial approximative
    network obtained from a small set of labeled training pairs, we
    estimate provisional model parameters for each unlabeled training
    exemplar. Given this provisional reflectance estimate, we then
    synthesize a novel temporary \emph{labeled} training pair by
    rendering the exact corresponding image under a new lighting
    condition. After refining the network using these additional
    training samples, we re-estimate the provisional model parameters
    for the unlabeled data and repeat the self-augmentation process
    until convergence.  We demonstrate the efficacy of the proposed
    network structure on spatially varying wood, metals, and plastics,
    as well as thoroughly validate the effectiveness of the
    self-augmentation training process.
    
\end{abstract}

%
%

%
%


\begin{teaserfigure}
  \centering
  \includegraphics[width=0.98\textwidth]{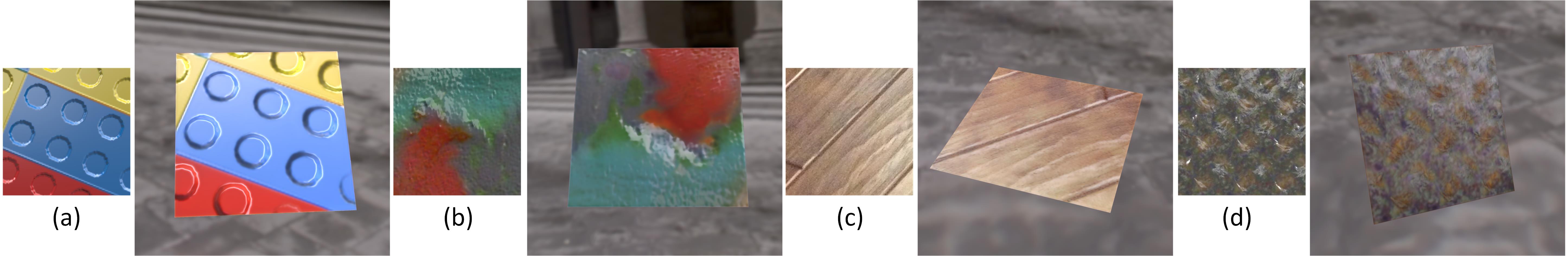}
  \caption{Physically plausible spatially varying surface appearance
    estimated using the proposed SA-SVBRDF-net from a single
    photograph of planar spatially varying \emph{plastic} (a,b),
    \emph{wood} (c) and \emph{metal} (d) captured unknown natural
    lighting, and revisualized under a novel lighting condition.  }
\label{fig:plausible}
\end{teaserfigure}

\maketitle

\section{Introduction}
\label{sec:intro}

Recovering the spatially varying bidirectional surface reflectance
distribution function (SVBRDF) from a single photograph under unknown
natural lighting is a challenging and ill-posed problem. Often a
physically accurate estimate is not necessary, and for many
applications, such as large scale content creation for virtual worlds
and computer games, a physically plausible estimate would already be
valuable.  Currently, common practice, albeit very time-consuming, is
to rely on skilled artists to, given a single reference image, produce
a plausible reflectance decomposition. This manual process suggests
that given sufficient prior knowledge, it is possible to infer a
plausible reflectance estimate for a spatially varying material from a
single photograph.

Data-driven machine learning techniques have been successfully applied
to a wide range of underconstrained computer graphics and computer
vision problems. In this paper, we follow a similar route and design a
Convolutional Neural Network (CNN) to estimate physically plausible
SVBRDFs from a single near-field observation of a planar sample of a
spatially varying material under unknown natural illumination.
However, recovering the SVBRDF from a single photograph is an
inherently ill-conditioned problem, since it is unlikely that each
pixel observes a significant specular response, making it impossible
to derive a full spatially varying specular component without
enforcing spatial priors.  We therefore estimate a \emph{reduced}
SVBRDF defined by a spatially varying diffuse albedo, homogeneous
specular albedo and roughness, and spatially varying surface normals.

Training a CNN to estimate such a reduced SVBRDF from a single
photograph under unknown natural lighting requires a large set of
``labeled'' photographs, i.e., with corresponding reflectance
parameters.  Gathering such a training dataset is often a tedious and
arduous task.  Currently, except for specialized materials, very few
databases exist that densely cover all possible spatial variations of
a material class.  Unlabeled data (i.e., a photograph of a spatially
varying material) is typically much easier to obtain. Each unlabeled
photograph contains an instance of the complex spatially varying
reflectance parameters, albeit it observed from a single view and
under unknown lighting.  This raises the question whether we can
exploit this embedded knowledge of the spatially varying distributions
to refine the desired SVBRDF-estimation CNN.

We propose, in addition to a CNN-based solution for SVBRDF estimation
from a single photograph, a novel training strategy to leverage a
large collection of \emph{unlabeled} data --photographs of spatially
varying materials without corresponding reflectance parameters-- to
augment the training of a CNN from a much smaller set of labeled
training data. To ``upgrade'' such unlabeled data for training, a
prediction of the unknown model parameters is needed. We propose to
use the target CNN itself to generate a provisional estimate of the
reflectance properties in the unlabeled photographs. However, we
cannot directly use the provisional estimates and the corresponding
unlabeled photograph as a valid training pair, since the estimated
parameters are likely biased by the errors in the CNN, and hence it
misses the necessary information to correct these errors.  Our key
observation is that for SVBRDF estimation, the \emph{exact} inverse of
the desired CNN is actually known in the form of a physically-based
rendering algorithm, that given any lighting and view parameters,
synthesizes a photograph of the estimated reflectance parameters.
Under the assumption that the initial CNN trained by the labeled data
acts as a reasonable predictor, the resulting provisional reflectance
estimates represent reasonable SVBRDFs similar (but not identical) to
the SVBRDFs in the unlabeled training photographs.  Therefore, instead
of directly using the provisional reflectance estimates and unlabeled
photographs as training pairs, we synthesize a new training sample by
rendering an image with the provisional reflectance estimates under
random lighting and view.  After refining the CNN using this
synthesized training data, we can update the provisional reflectance
estimates and corresponding synthetic visualizations, and repeat the
process.  The proposed self-augmentation training process
progressively refines the CNN to be coherent with the known inverse
process (i.e., rendering algorithm), thereby improving the accuracy of
the target CNN.  We demonstrate the efficacy of our method by training
a CNN for different classes of spatially varying materials such as
wood, plastics and metals, as well as perform a careful analysis and
validation of the proposed self-augmentation training strategy.

\section{Related Work}
\label{sec:related}

\paragraph{Single Image Reflectance Modeling}
%
For conciseness, we focus this overview on single image reflectance
modeling; for an in-depth compilation of general data-driven
reflectance estimation techniques we refer to the excellent overviews
by Dorsey et al.~\shortcite{Dorsey:2007:DMM} or Weinmann and
Klein~\shortcite{Weinmann:2015:AGR}.

In general, estimation of surface reflectance from a single photograph
is an ill-posed problem.  A common strategy to make estimation more
tractable is to control the incident lighting during acquisition,
often by limiting the lighting to a single directional light
source. Wang et al.~\shortcite{Wang:2016:SIS} recover shape, spatially
varying diffuse albedo and a homogeneous specular component from a
light field observation under a single known directional light
source. Xu et al.~\shortcite{Xu:2016:MBS} describe a general framework
for recovering (piecewise constant) homogeneous surface reflectance
from near-field observations lit by a known directional light source.
Aittala et al.~\shortcite{Aittala:2016:RMN} model the spatially
varying surface reflectance and surface normals from a single flash
image of a stationary textured material. They avoid the need for
explicit point-to-point surface correspondences by relying on a
powerful CNN-based texture descriptor for assessing the quality of the
predictions. The Deep Lambertian Network~\cite{Tang:2012:DLN} is a
full end-to-end machine learning approach, based on belief networks,
for jointly estimating diffuse reflectance, surface normals, and the
lighting direction. Finally, in contrast to the previous methods that
rely on directional light sources, Wang et
al.~\shortcite{Wang:2011:EDP} recover the surface reflectance and
complex spatially varying surface normals of a homogeneous glossy
material using step-edge lighting.  All these methods rely on active
illumination which limits their practical use to fully controlled
settings or environments with minimal ambient lighting.  The proposed
method does not rely on active illumination, and it does not make any
assumptions on the form of the incident lighting.

Oxholm and Nishino~\shortcite{Oxholm:2012:SRN,Oxholm:2016:SRE} relax
the requirement of active illumination and recover the shape and
homogeneous surface reflectance from a single photograph under
uncontrolled, but \emph{known}, lighting.  Similarly, Hertzmann and
Seitz~\shortcite{Hertzmann:2003:SAM} recover spatially varying surface
reflectance and surface normals under uncontrolled lighting given a
reflectance map of each material (i.e., a photograph of a spherical
exemplar under the target lighting).  While these methods recover the
surface reflectance from a single photograph, they do require
additional measurements to obtain the required lighting/reflectance
reference maps.  

Romeiro and Zickler~\shortcite{Romeiro:2010:BR} assume the incident
lighting is \emph{unknown}, and propose to recover the most likely
reflectance of a spherical homogeneous object under the expected
natural lighting distribution. Similarly, Lombardi and
Nishino~\shortcite{Lombardi:2012:RNI,Lombardi:2016:RIR} express the
reflectance recovery of a homogeneous object under unknown lighting as
a maximum a-posteriori estimation with strong priors on both lighting
and surface reflectance.  Barron et al.~\shortcite{Barron:2015:SIR}
propose to find the most likely shape, piecewise constant spatially
varying diffuse reflectance, and illumination that explains the input
photograph based on strong hand-crafted priors on each
component. Rematas et al.~\shortcite{Rematas:2016:DRM} employ a
convolutional neural network to extract homogeneous \emph{reflectance
  maps} under unknown lighting and shape.  All these methods are
limited to either diffuse or homogeneous materials. The proposed
method on the other hand, is specifically designed for spatially
varying materials.

Finally, AppGen~\cite{Dong:2011:AIM} takes a different approach to
separate the different reflectance components from a single photograph
by putting the user in the loop. While AppGen greatly accelerates the
authoring process, it does not scale well to large-scale content
generation due to the required (albeit limited) manual
interaction. Furthermore, AppGen assumes the input image is
illuminated by a single directional light.

\paragraph{Deep Learning with Unlabeled/Synthetic Data}
Recent advan\-ces in deep learning have been successfully applied to a
wide variety of computer graphics and computer vision problems. A full
overview falls outside the scope of this paper, and we focus our
discussion on prior work that combines deep learning techniques with
unlabeled and/or synthetic training data.

A common weakness for many deep learning methods is their reliance on
large training datasets which can be difficult and/or time-consuming
to obtain.  To alleviate the difficulty in gathering such datasets,
several researchers have looked at synthetic data to fill in the
gaps. Synthetic data has been used to improve the training of various
tasks, for example for: text detection~\cite{Gupta:2016:SDT}, hand
pose estimation~\cite{Tompson:2014:RCP}, object
detection~\cite{Gupta:2014:LRF}, semantic segmentation of urban
scenes~\cite{Ros:2016:TSD}, etc.  Gaidon et
al.~\shortcite{Gaidon:2016:VWP} showed that pretraining on synthetic
data improves the overall accuracy of the regression.  Narihira et
al.~\shortcite{Narihira:2015:DIL} propose ``Deep IntrinsicNet'', a CNN
for single image intrinsic decomposition; a subject closely related to
reflectance estimation, and which decomposes an image in its
``shading'' and ``reflectance'' components.  They observe that
marrying synthetic training data with CNNs is a powerful paradigm, and
follow Chen et al.~\shortcite{Chen:2013:ASM} to train their
IntrinsicNet using the MPI Sintel dataset. We also rely on synthetic
training data, but instead of only synthesizing training data, we also
use it to complete non-synthetic unlabeled training data.

Shelhamer et al.~\shortcite{Shelhamer:2015:SIA} also rely on CNNs to
infer the intrinsics from a single image. However, they note that CNNs
cannot guarantee physical coherence, and propose to combine and
jointly train a CNN for surface normal estimation together with a
fixed inverse rendering pipeline to compute the intrinsic
decomposition (e.g., using the method of Barron et
al.~\shortcite{Barron:2015:SIR}).  The proposed method also includes a
rendering component, but instead of appending an inverse rendering
pipeline to the CNN, we employ a \emph{forward} rendering pipeline to
aid in training by synthesizing new training pairs from unlabeled
data.

Breeder Learning~\cite{Nair:2008:ASL} relies on a known generative
black-box function to generate novel training data by perturbing the
estimated parameters, and computing the resulting exemplar using the
black-box generator.  In contrast, self-augmentation (where the
forward rendering component could be seen as a generative black-box)
relies on unlabeled data instead of perturbations to explore and
refine the search space. Furthermore, perturbations in the parameter
space do not necessarily match the real-world distribution of the
underlying data, especially for high-dimensional parameter spaces
(e.g., a random perturbation on the diffuse albedo or on the surface
normals does not necessarily yield a valid SVBRDF within the targeted
material class).

\section{SVBRDF-Net}
\label{sec:svbrdf}

\begin{figure}[t]
  \begin{center}
  \includegraphics[width=0.48\textwidth]{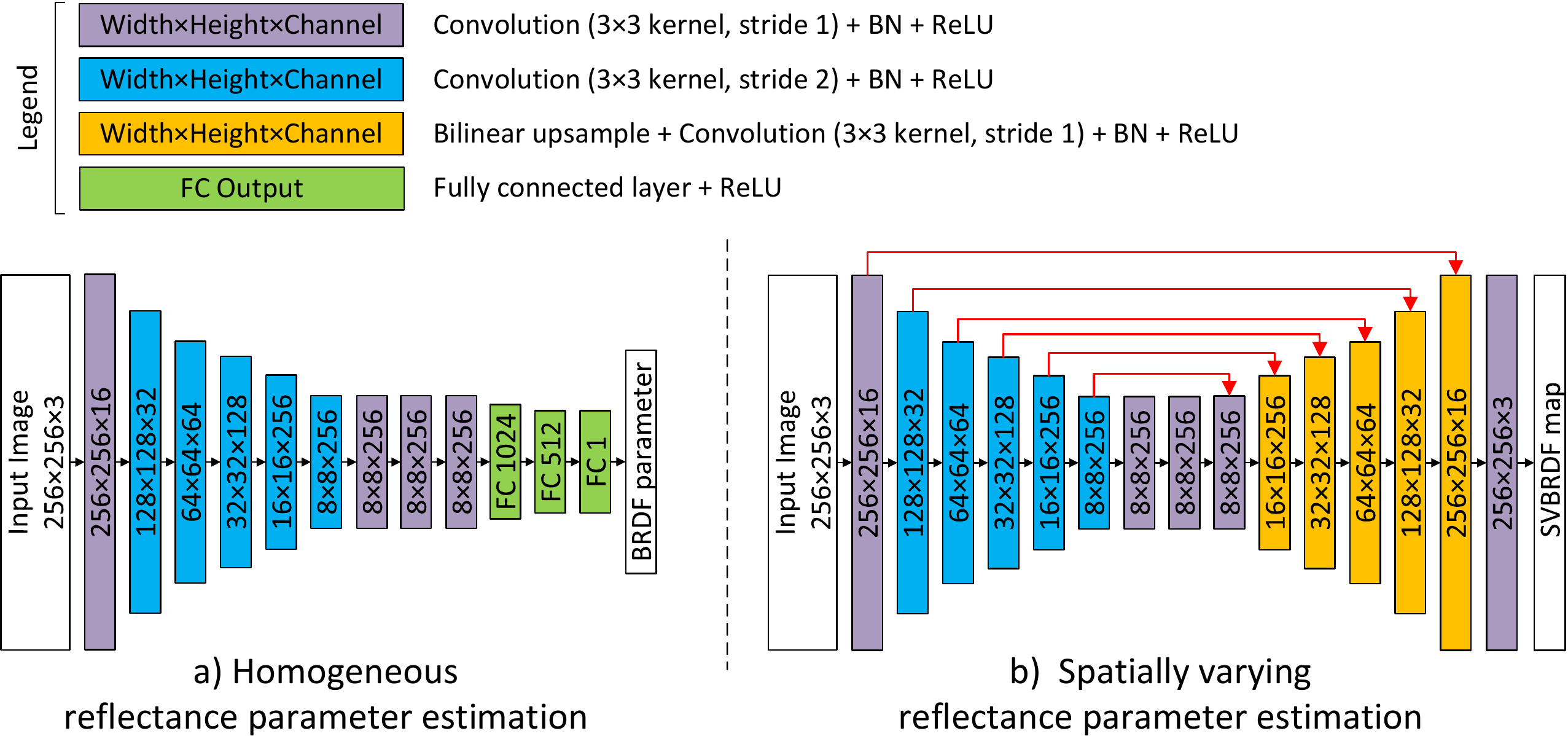}
  \end{center}
  \caption{ \textbf{Network structure.} Left: the network structure
    for homogeneous reflectance parameters consists of an analysis
    subnetwork of convolution and pooling layers, followed by a fully
    connected synthesis network with $1024$ hidden variables.  Right:
    the network for spatially varying reflectance parameters consists
    of an identical analysis network of convolution and pooling
    layers, followed by a synthesis upsampling network that mirrors
    the analysis network. To reintroduce high frequency details during
    upsampling, the feature maps from the analysis layers are
    concatenated to the corresponding synthesis layers (illustrated by
    the red links). }
\label{fig:network}
\end{figure}

\paragraph{Convolutional Neural Network Configuration}
Our goal is to estimate plausible appearance parameters from a single
photograph of a near-field observation of a planar material sample
under unknown natural lighting.  Modeling both spatially varying
diffuse as well as specular reflectance from a single photograph is an
ill-posed problem. We therefore reduce the complexity by assuming a
homogeneous constant specular component on top of a spatially varying
diffuse component and surface normals.  We introduce a convolutional
neural network solution (SVBRDF-net) to learn the mapping from a
photograph of a spatially varying material to appearance parameters,
where the homogeneous specular component is represented by the Ward
BRDF model~\cite{Ward:1992:MMA} parameterized by its specular albedo
$\ks$ and specular roughness parameter $\roughness$, the spatially
varying diffuse component is modeled by a per-pixel diffuse albedo
$\kd(\pos)$, and the surface normals are encoded as a 3D vector per
pixel $\n(\pos)$:
\begin{equation}
  \fr(\wi, \wo, \pos) = \frac{\kd(\pos)}{\pi} + \ks \frac{\exp{-\tan^2\delta / \roughness^2}}{4\pi\roughness^2 \sqrt{\dot{\wi}{\n(\pos)}\dot{\wo}{\n(\pos)}}},
\end{equation}
where $\delta$ the angle between the halfway vector $\wh = (\wi + \wo)
/ || \wi + \wo ||$ and the normal $\n(\pos)$.  Note that the proposed
SVBRDF-net is not married to the Ward BRDF model and any other BRDF
model can be used instead.

Due to the inherent scale-ambiguity between the overall intensity of
the lighting and surface reflectance (i.e., scaling the intensity of
the lighting can be compensated by dividing the diffuse and specular
albedo by the same factor), SVBRDF-net can only estimate relative
diffuse albedos $\kd_{rel}$ and the relative specular albedo
$\ks_{rel}$. We will assume without loss of generality that the
average diffuse albedo (over all color channels) is fixed ($avg(\kd) =
0.5$), and that the relative diffuse and specular albedo are expressed
with respect to this average diffuse albedo: $\kd_{rel}(\pos) =
\kd(\pos) / s$, $\ks_{rel} = \ks / s$ and $s = 2
avg(\kd)$. Consequently, any uniform scaling applied to $\kd$ and
$\ks$ does not affect $\kd_{rel}$ and $\ks_{rel}$. For stability, we
use the logarithm of the relative specular albedo ($\ks_{log-rel} =
\log{\ks_{rel}}$) to avoid huge parameter values for the specular
component in SVBRDF-net when the absolute average diffuse albedo ($s$)
is small.  Furthermore, to ensure a more uniform distribution with
respect to the error on the appearance, a log-transformation is also
applied to the specular roughness encoded in SVBRDF-net:
$\roughness_{log} = \log{\roughness}$.

The proposed SVBRDF-net is a union of separate network structures; one
for the homogeneous parameters (relative (log) specular albedo and
(log) roughness), and another for the spatially varying parameters
(relative diffuse albedo and surface normals). \autoref{fig:network}
depicts the network structure and lists the relevant dimensions of the
layers.  The number of layers and convolution/upsample filter sizes
are similar to those used in prior work
(e.g.,~\cite{Rematas:2016:DRM}). Both networks share the same analysis
subnetwork structure consisting of a series of convolution layers and
pooling layers. Each convolution layer is followed by a
batch-normalization layer~\cite{Ioffe:2015:BN} and a ReLU activation
layer. This analysis subnetwork reduces the input photograph to the
essence needed for estimating the relevant model parameters. The
analysis subnetwork is followed by a different synthesis subnetwork
for each output parameter type:
\begin{itemize}[leftmargin=0.175in]
  \myitem{Homogeneous Specular Albedo and Roughness:} are
    synthesized by adding a fully connected layer with 1024 hidden
    variables, and $6$ output nodes (i.e., relative (log) specular
    albedo $\ks_{log-rel}$ and (log) roughness $\roughness_{log}$
    estimates per color channel).
  \myitem{Spatially Varying Diffuse Albedo:} is synthesized by a
    set of upsample+convolution layers (often misnamed as a
    ``deconvolution'' layer). Similarly as for analysis, each layer is
    followed by a batch normalization and ReLU layer before a bilinear
    upsampling step. As
    in~\cite{Rematas:2016:DRM,Ronneberger:2015:UCN}, we concatenate
    feature maps from the corresponding analysis layers (marked by the
    red links in~\autoref{fig:network}) to help reintroduce high
    frequency details lost in subsequent convolution layers during
    analysis. The final output is mapped to the proper range (i.e.,
    $[0,1]$) using a standard per-pixel fixed sigmoid function. As batch
    normalization does not work well with a sigmoid output layer, we
    omit the batch normalization at the final output layer.
    \myitem{Spatially Varying Normal Map:} is synthesized by an
      identical (but separate) network as for the diffuse
      albedo.
\end{itemize}
We train a separate analysis-synthesis network for each output
parameter type.

\paragraph{Training by Self-Augmentation}
One of the main challenges in training the proposed SVBRDF-net, is to
obtain a sufficiently large training dataset that captures the natural
distribution of all spatial variations for the target material
class. For each spatial variation, we need, in addition, a sufficient
sampling of the different viewing and lighting
conditions. Essentially, the search space we are regressing is the
outer product of the appearance variations due to spatial variations
in reflectance properties, different natural lighting conditions, and
differences in viewpoint.  The latter two dimensions are common over
different material classes and are relatively easily to sample. The
former, on the other hand, is highly dependent on the material type
and is ideally sampled through diversity in the training data.

The desired labeled training data in the case of SVBRDF-net is in the
form of a photograph of a planar sample of an spatially varying
material under an unknown natural lighting condition paired with the
corresponding model parameters (i.e., spatially varying relative
albedos, log-roughness, and surface normals). Note that the high cost
of obtaining labeled training data is solely due to the cost of the
necessary acquisition/authoring process to obtain the model
parameters.  On the other hand, the cost to obtain unlabeled data in
the form of a photograph of a planar spatially varying material is
significantly smaller.  Similarly to labeled data, such unlabeled data
represents samples in the search space, except that the corresponding
model parameters are unknown. Now suppose we have an oracle that can
predict these model parameters from the unlabeled data, then we can
use the unlabeled data in conjunction with the prediction to refine
SVBRDF-net.  Of course, this oracle is exactly the network we desire
to train.  Now let's assume we have a partially converged SVBRDF-net,
trained from a smaller set of labeled training data.  This partially
converged network forms an approximation of the desired oracle.
Hence, we can use the network itself to generate provisional model
parameters for each unlabeled exemplar. However, we cannot directly
use these provisional model parameters paired with the input unlabeled
image for training since the provisional model parameters are biased
by the errors in the partially converged network (and thus cannot
correct these errors).

\begin{figure*}[th!]
  \begin{center}
  \includegraphics[width=0.98\textwidth]{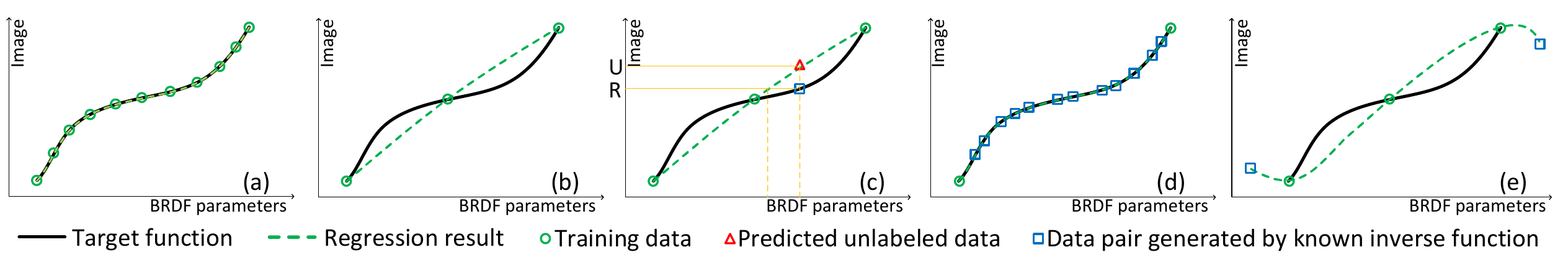}
  \end{center}
  \vspace{-0.2cm}
  \caption{ \textbf{Illustration of the Self-Augmentation Training
      Process}.  (a) Given sufficiently dense sampled training data,
    the target manifold can be well modeled. (b) However, when
    insufficiently sampled, the target network only provides a coarse
    approximation. (c) Self-augmentation exploits the fact that a
    forward rendering algorithm provides an \emph{exact} inverse of
    the target manifold. Given a coarse approximation of the manifold,
    we can make a provisional estimate of the reflectance parameters
    of an unlabeled image (marked with 'U'). Rendering the estimated
    provisional parameters from a new viewpoint and under novel
    lighting, produces a different image ('R') than the unlabeled
    image. However, unlike the unlabeled image, this rendered image is
    not biased by the approximation error of the coarse network. When
    inputting this rendered image to the target CNN, we obtain a
    different estimate of the parameters than the ones used for
    rendering. Self-augmentation attempts to reduce the mismatch
    between the provisional and estimated parameters from the rendered
    image, thereby indirectly improving the accuracy of the target
    network. (d) Repeatedly applying the self-augmentation process on
    a large set of unlabeled data points, should yield a progressively
    more accurate approximation of the target network. (e) However,
    care must be taken that the unlabeled data points lie in or near
    the search space covered by the labeled data points as the latter
    determines the range in which the initial coarse approximative
    network is well-defined.  }
\label{fig:illustration}
\end{figure*}

\begin{figure}[th!]
  \begin{center}
  \includegraphics[width=0.45\textwidth]{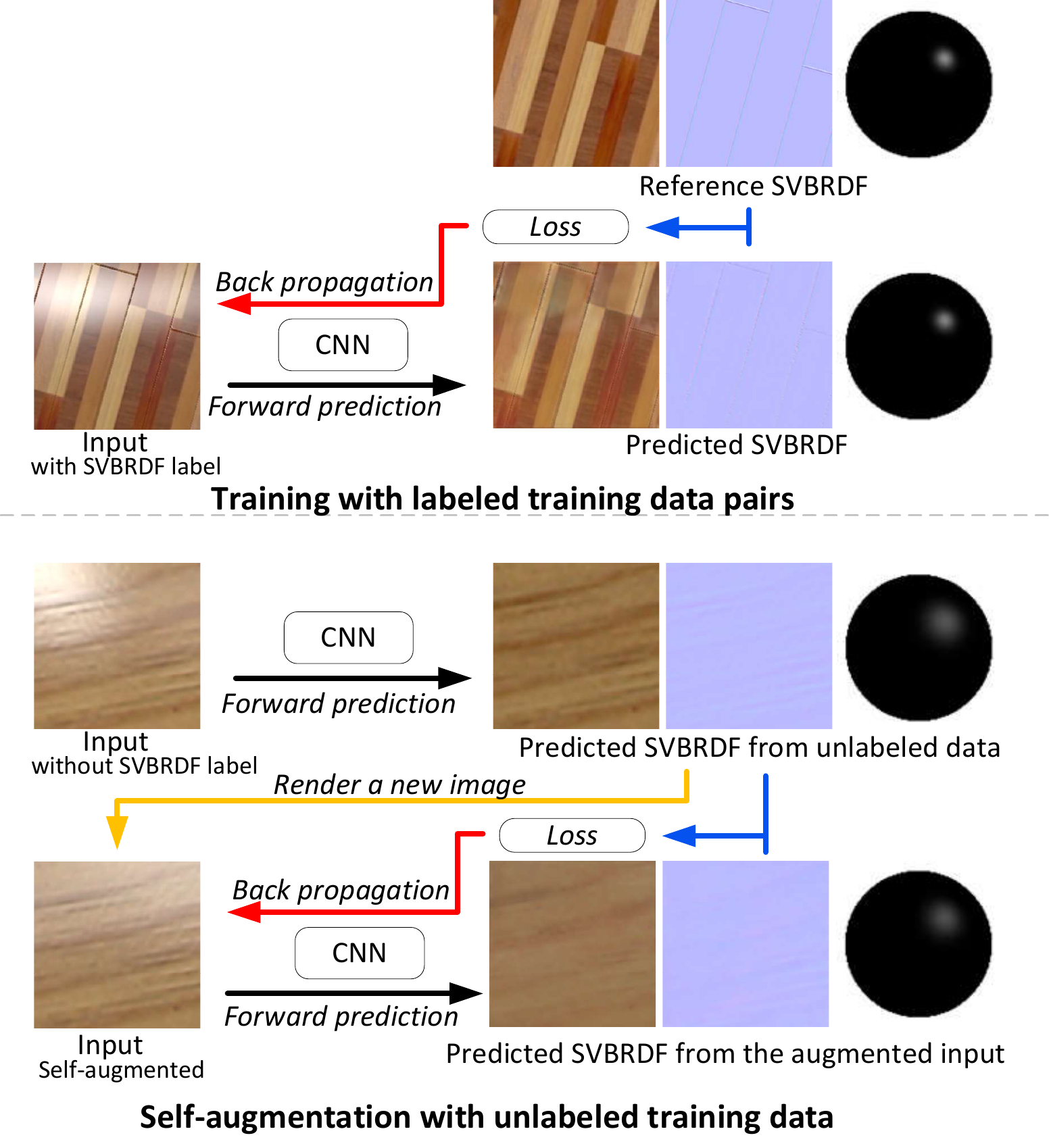}
  \end{center}
  \vspace{-0.1cm}
  \caption{\textbf{Summary of the Self-augmentation Training Process.}
    First, an initial approximative network is trained using only
    labeled data. Next, we alternate between training on labeled and
    unlabeled data. For each unlabeled exemplar, we estimate
    provisional reflectance parameters, and synthesize a new
    (temporary) labeled training pair by rendering a corresponding
    \emph{exact} visualization. This new labeled training pair, is
    then used to further refine the network.}
\vspace{-0.4cm}
\label{fig:loop}
\end{figure}

Our key observation is that the \emph{exact} inverse of the target
network is known in the form of a rendering algorithm, i.e., given the
lighting and view parameters together with the model parameters, we
can synthesize a photograph of the SVBRDF.  Our goal is to exploit
this (inverse) knowledge of the search space to refine the regression
of the CNN.  Assuming a locally smooth search space, and assuming the
unlabeled data lies in the search space covered by the labeled
training data (i.e., the labeled data covers the full target search
space), then the provisional model parameters should approximately
follow the target distribution. Hence the extracted spatially varying
model parameters form a plausible SVBRDF (albeit not necessarily the
same as the ground truth physical parameters of the unlabeled data
exemplar).  Thus, the provisional model parameters and a corresponding
synthesized visualization under either the same or different lighting
and view conditions, provide a valid labeled training exemplar that
can be used to refine the CNN. This process is illustrated
in~\autoref{fig:illustration}.  Given sufficient training data (a), we
can accurately capture the target search space's manifold. However,
when a smaller labeled training set is used (b), the resulting
regression only forms a coarse approximation. Given unlabeled data
(marked by 'U' on the vertical axis in (c)), we can estimate
provisional model parameters using the initial network as an oracle
(the corresponding projection on the CNN is marked by a triangle).
The rendering process takes these provisional model parameters and
'projects' them back to image space over the exact manifold, resulting
in a different rendered image (marked by 'R' on the vertical axis).
When using this synthetic pair for training, we observe a discrepancy
between the estimated model parameters from this synthetic image
(i.e., the projection of the synthetic rendering ('R') over the CNN
(dashed line)) and the initial provisional model parameters. Hence,
the training process will minimize this discrepancy. Assuming local
smoothness over the search space and assuming that the initial CNN
prediction is relatively close to the target manifold, minimizing this
discrepancy should locally pull the CNN's manifold approximation
closer to the desired search space's manifold (d).  However, care must
be taken to ensure that the unlabeled data resides in or near the
search space covered by the labeled training data (e), as there is no
guarantee about the extrapolation behavior of the CNN outside this
region, and thus the provisional model parameters are unlikely to
represent a plausible SVBRDF in the targeted material class and/or
potentially correspond to a folding of the manifold (thereby creating
ambiguities in the training).

We coin our training strategy ``self-augmenting'' since it relies on
the exactness of the inverse model (i.e., the rendering algorithm) to
guide the training, and the CNN itself to provide reasonable model
parameters. We integrate this self-augmentation in the training
process, and repeatedly apply it on the progressively refined network
(and thus with progressively improved provisional model
parameters). We will use the short-hand notation SA-SVBRDF-net to
refer to the self-augmented SVBRDF-net, and use SVBRDF-net to refer to
the regularly trained SVBRDF-net (i.e., without
self-augmentation). ~\autoref{fig:loop} summarizes the proposed
self-augmentation training strategy.


\section{Results}
\label{sec:results}

\paragraph{Implementation}
We implement the proposed SVBRDF-net detailed in~\autoref{sec:svbrdf}
in Caffe~\cite{Jia:2014:Caffe} using a constant initialization
strategy and train it with ADAM~\cite{Kingma:2015:AMS}. We first
produce an initial CNN by training, from scratch, with labeled data
only for $10$ epochs. Next, we self-augment this initial SVBRDF-net
using the unlabeled training data.  However, whereas labeled data
provides absolute cues to the structure of the search space, the
unlabeled data only constrains the local structure of the search space
(i.e., it ensures that the network is locally consistent with the
rendering process). Therefore, care should be taken to ensure that the
self-augmentation does not steer the network too far from the labeled
data, and avoid collapsing the network to a (self-coherent) singular
point.  Hence, we interleave the training after each randomly selected
unlabeled mini-batch, with a random labeled mini-batch that biases the
training to remain faithful to the labeled
data. \autoref{tab:parameters} summarizes all relevant training
parameters. For all results in this paper, we trained the proposed
SA-SVBRDF-net for an input resolution of $256 \times 256$.

\begin{table}[t]
\caption{\textbf{Summary of Training Hyperparameters}. We use the
  default Caffe values for unlisted parameters.}
\vspace{-0.3cm}
\small
\begin{center}
\begin{tabular}{|c|c|c|c|c|c|}
\hline
\multicolumn{3}{|c|}{Learning rate}&Weight&Momentum&Mini-batch \\
\cline{1-3}
Inital & Policy & Gamma &  decay &  &size \\
\hline
0.002 & Inverse & $10^{-4}$ & $10^{-4}$ & 0.9 & 16\\
\hline
\end{tabular}
\end{center}
\label{tab:parameters}
\end{table}

\begin{figure*}
  \begin{center}
  \includegraphics[width=.92\textwidth]{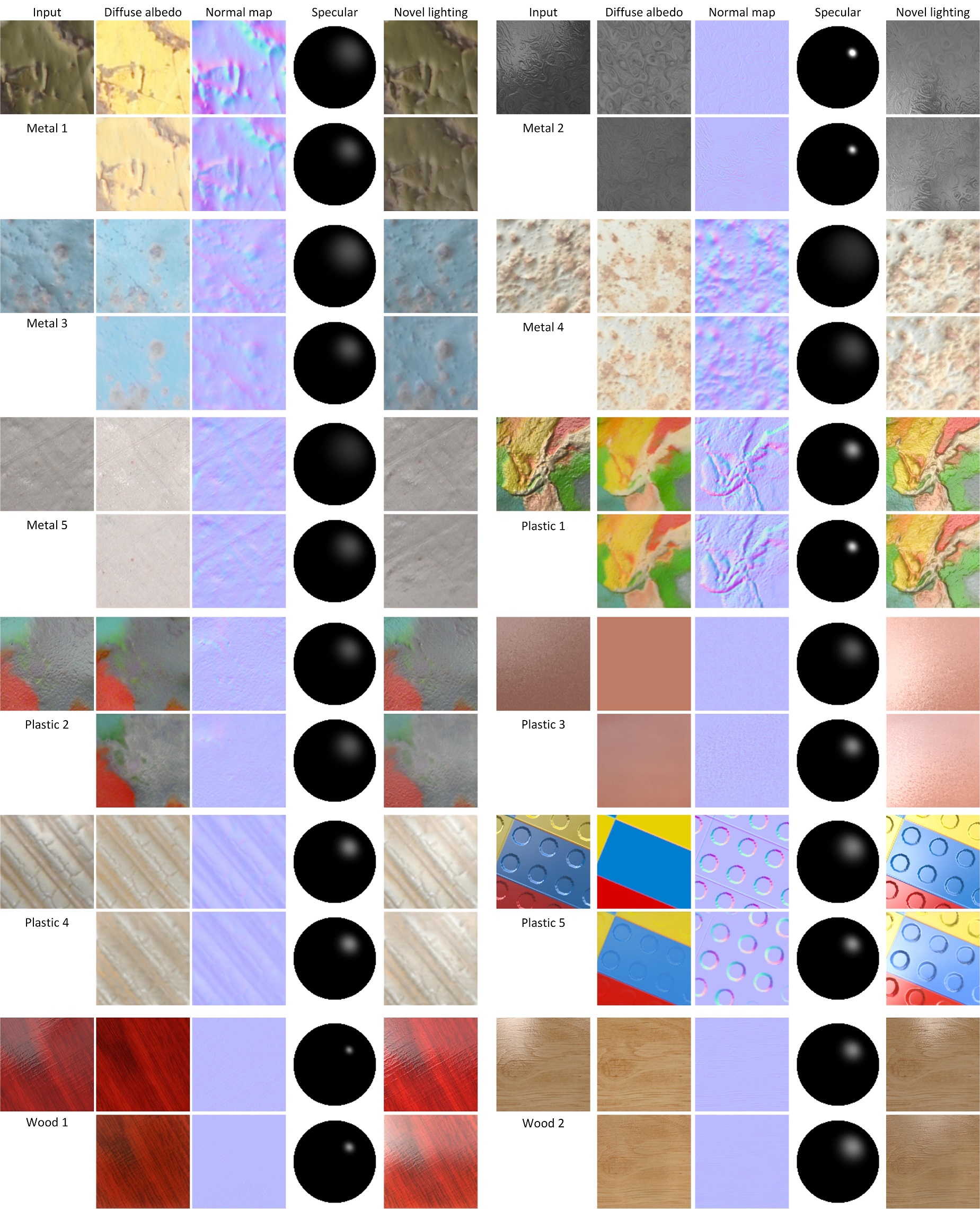}
  \end{center}
  \caption{\textbf{Qualitative Comparison to Reference Reflectance
      Parameters.} Reflectance parameters obtained with the
    \emph{wood}, \emph{plastic}, and \emph{metal} SA-SVBRDF-nets (odd
    rows) compared to reference reflectance parameters (even
    rows). Each row shows (left to right), spatially varying relative
    diffuse albedo, normal map, the homogeneous relative specular
    albedo and roughness mapped on a sphere and illuminated by a
    directional light source, and a visualization of the
    reference/estimated appearance parameters under a novel lighting
    condition.}
\label{fig:results}
\end{figure*}

\paragraph{Data Collection \& Preprocessing}
We source SVBRDF-data for spatially varying \emph{wood},
\emph{plastics}, and \emph{metals} from an online material
library~\cite{materials}, supplemented by artist generated SVBRDFs. We
average specular albedo and roughness when the dataset provides
spatially varying specular albedo or roughness.  The \emph{wood}
dataset contains $50$ SVBRDFs, the \emph{plastic} dataset contains
$60$ SVBRDFs and the \emph{metal} dataset contains $60$ SVBRDFs.  We
randomly select $10$ SVBRDFs from each dataset for testing, and retain
the remainder for training (thus $40$ for \emph{wood}, and $50$ for
\emph{plastic} and \emph{metal}). To generate labeled training/test
data, we furthermore gather $50$ HDR light probes~\cite{hdrimages},
and render (with a GPU rendering implementation that correctly
integrates the lighting from the full hemisphere of incident
directions) each SVBRDF under $40$ selected light probes for the
training set, and the remaining $10$ probes for the test set, for $9$
randomly selected rotations. Consequently, the lighting for each
labeled training pair, as well as the test set, is different. During
training, we randomly crop and rotate a $256 \times 256$ window from
the rendered images to sufficiently sample the rich texture
variations.  As we aim for a practical tool, we only use
radiometrically linearized ``low'' dynamic range images (i.e., all
pixel values are clamped to the $[0, 1]$ range). If the input image is
not radiometrically linear, we apply a gamma (2.2) correction before
feeding it to the proposed SA-SVBRDF-net.

Unlabeled data is gathered from OpenSurface~\cite{Bell:2013:ORA}, as
well as internet images collections.  In total, the number of
unlabeled images are $1000$, $1000$, and $1200$ for \emph{wood},
\emph{plastic} and \emph{metal} SA-SVBRDF-net training
respectively. During self-augmentation, we randomly pick and rotate a
lighting environment for generating the provisional labeled training
pair using the same rendering algorithm as was used for generating the
training data.  We assume that both the labeled and unlabeled training
data are drawn from the same distribution. However, both (labeled and
unlabeled) databases are likely to be gathered from different sources
or created through a different authoring process. Consequently, their
respective inherent distribution through the search space might
differ. To avoid biasing the training, we select a subset from the
unlabeled data such that it mimics the distribution of the labeled
data.  However, we do not know the reflectance parameters for the
unlabeled data, and thus also not their distribution.  Instead, of
matching the search space distribution, we match the first order
statistics of the images as a proxy.  Practically, we create the
histogram of the average colors of the labeled training images, and
then randomly select the subset of the unlabeled images such that the
resulting histogram follows the same distribution. While a crude
approximation, we have found this to work well in practice.
Furthermore, we auto-expose the rendered synthetic images before
clamping pixel values to the $[0,1]$ range. This allows us to directly
use the relative albedos during rendering instead of converting them
to absolute parameters; any scale applied to both the relative diffuse
and specular albedo is compensated for by a corresponding change in
the auto-exposure.

As noted in~\autoref{sec:svbrdf}, to resolve the ambiguity between the
intensity of the lighting and the reflectivity of the surface, we
ensure that the average diffuse albedo over all color channels equals
0.5, and express the specular albedo with respect to this
average. However, a related ambiguity still exists between the color
of the lighting and the material.  We cannot simply set the average
diffuse albedo for each color channel to $0.5$ as this would result in
a complete loss of material color.  Instead, when generating labeled
training samples, we white-balance the lighting such that the
irradiance is color neutral:
\begin{equation}
  \forall i,j \in [r, g, b]: s_i = s_j, ~\text{and}~~ s_{r,g,b} = \int_{\Omega_+} \E(\wi) \cos{\theta_i} d\wi,
\end{equation}
where $\E(\wi)$ is the incident lighting.  In addition, we also assume
that unlabeled data and/or input photographs are (approximately)
correctly white balanced. In other words, we assume that the diffusely
reflected color observed in the input photograph has the same color as
the diffuse albedo.

\paragraph{Results}
\autoref{fig:plausible} shows a series of input photographs for the
\emph{wood}, \emph{plastic}, and \emph{metal} SA-SVBRDF-nets and a
revisualization under a novel lighting condition for the recovered
spatially varying reflectance parameters. Qualitatively, we argue that
the appearance of the revisualizations exhibits the same visual
qualities as the input photographs. This suggests that the proposed
SA-SVBRDF-nets are able to estimate plausible reflectance parameters.

While our goal is to model plausible surface reflectance from a single
photograph, it is nevertheless informative to explore how well the
estimated parameters match ground truth surface reflectance
parameters.  \autoref{fig:results} shows a comparison between ground
truth reflectance parameters and the estimated parameters obtained
with the \emph{wood}, \emph{plastic}, and \emph{metal} SA-SVBRDF-nets.
For each example, we show two rows, where the top row shows the ground
truth and the bottom row shows the recovered results.  For each row,
we show (from left to right), the input photograph under an unknown
natural lighting condition, the recovered diffuse albedo, the
recovered normal map, the homogeneous specular component mapped on a
sphere and lit by a directional light, and a visualization under a
novel lighting condition.  Note that none of the input
photographs/SVBRDFs were included in the training set to avoid bias in
the results. While there are some differences, overall, the estimated
reflectance parameters match the reference parameters well.  This
further confirms that the resulting reflectance parameters estimated
with SA-SVBRDF-net are physically plausible.

\begin{figure}[t]
  \begin{center}
  \includegraphics[width=0.45\textwidth]{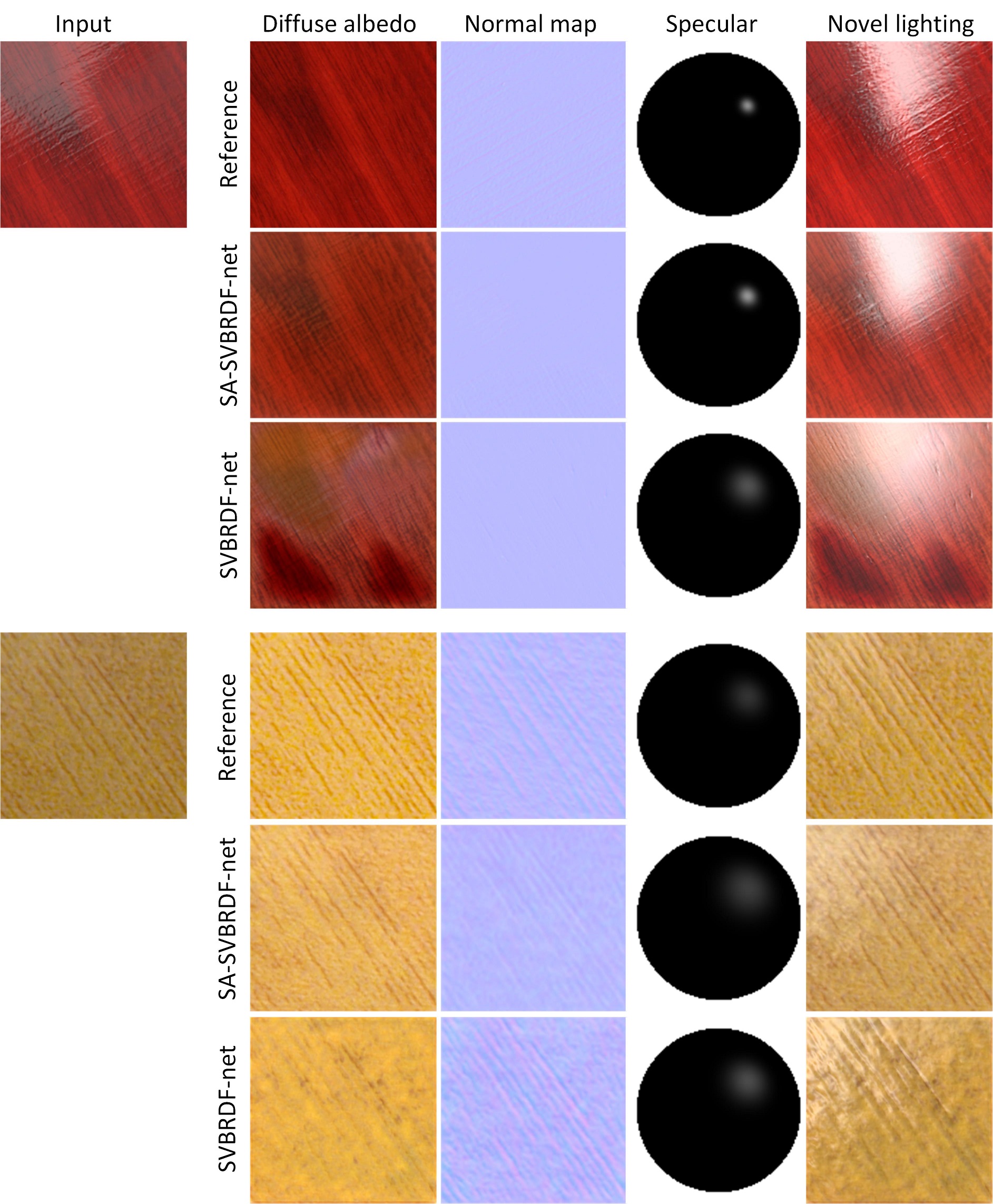}
  \end{center}
  \vspace{-0.2cm}
  \caption{\textbf{Impact of Self-Augmentation.} A qualitative
    comparison between reflectance parameters estimated with
    SVBRDF-net trained with and without self-augmentation. The
    SA-SVBRDF-net estimates appear more plausible and exhibit less
    artifacts than the results from the regular SVBRDF-net. }
\vspace{-0.2cm}
\label{fig:loop_comp}
\end{figure}

\begin{figure}[t]
  \centering
  \includegraphics[width=0.45\textwidth]{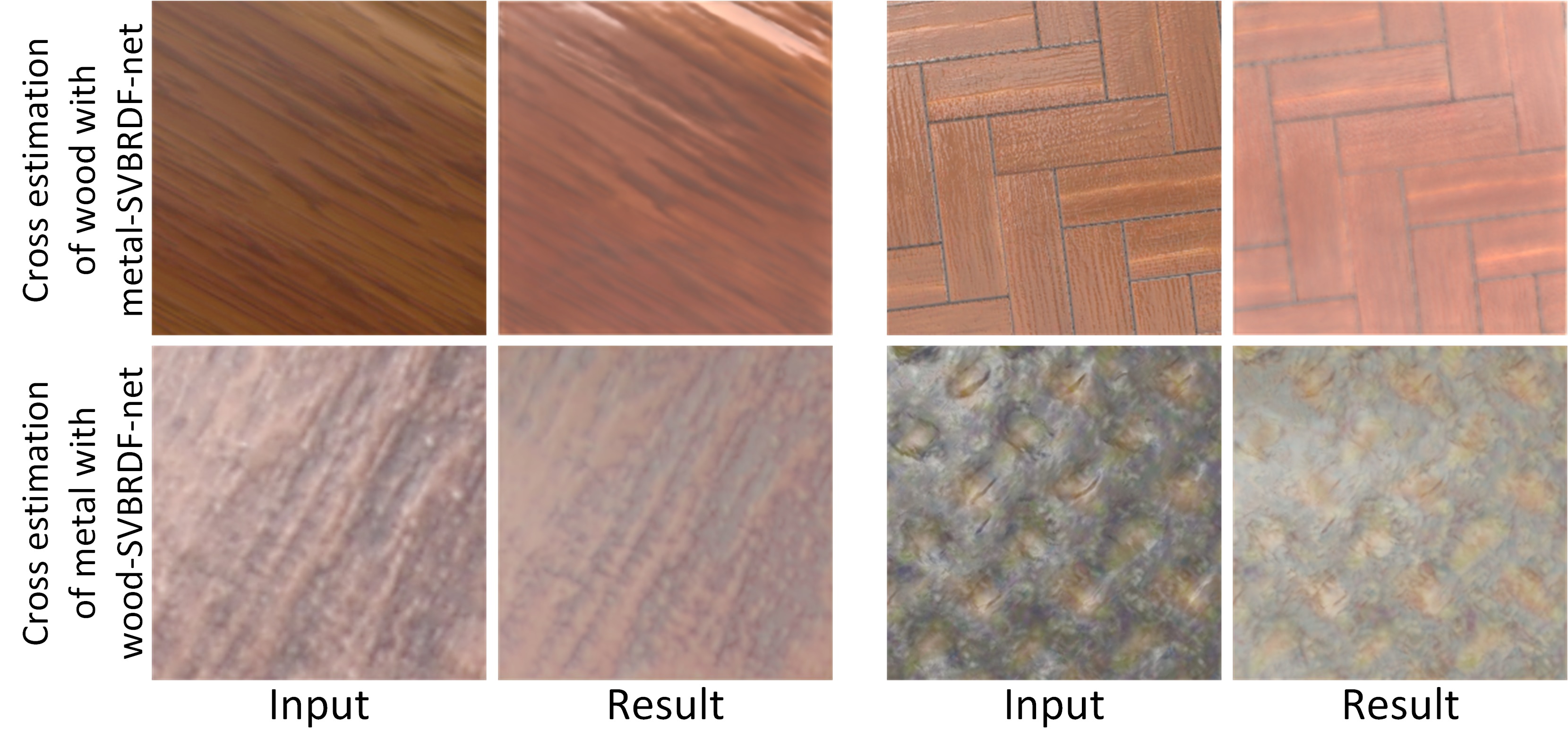}
  \vspace{-0.2cm}
  \caption{\textbf{Cross Estimation.} Surface appearance of
    \emph{wood} and \emph{metal} estimated using an SVBRDF-net trained
    on another material class (metal and wood respectively) fails to
    produce physically plausible results.}
  \vspace{-0.3cm}
  \label{fig:cross}
\end{figure}

\autoref{fig:loop_comp} compares the estimated reflectance estimates
of the \emph{wood} SVBRDF-net and SA-SVBRDF-net. In both cases, we use
exactly the same set of $40$ labeled training samples.  Compared to
the regular SVBRDF-net, the proposed SA-SVBRDF-net trained with
self-augmentation produces qualitatively more plausible results with
less visual artifacts.  For example, the top example includes remnants
of the specular highlight from the input photograph in the diffuse
albedo while overestimating the specular roughness, and the bottom
example exhibits a faint ``splotchy'' structure in the diffuse
albedo. This empirically shows that the proposed self-augmentation
strategy greatly helps the convergence of the training process, and
significantly reduces the required number of labeled training samples.
Both networks were trained on a NVidia Titan X (Maxwell) for $20$
epochs; $10$ epochs to obtain the rough initial SVBRDF-net, and then
an additional $10$ with-out/with unlabeled data for the SVBRDF-net and
SA-SVBRDF-net respectively. Total training time without
self-augmentation took $30$ hours, and $38$ hours with
self-augmentation.  Evaluating the (SA-)SVBRDF-net is very fast and
only takes $0.3$ seconds on a GPU.

\begin{figure}[t]
  \centering
  \includegraphics[width=0.45\textwidth]{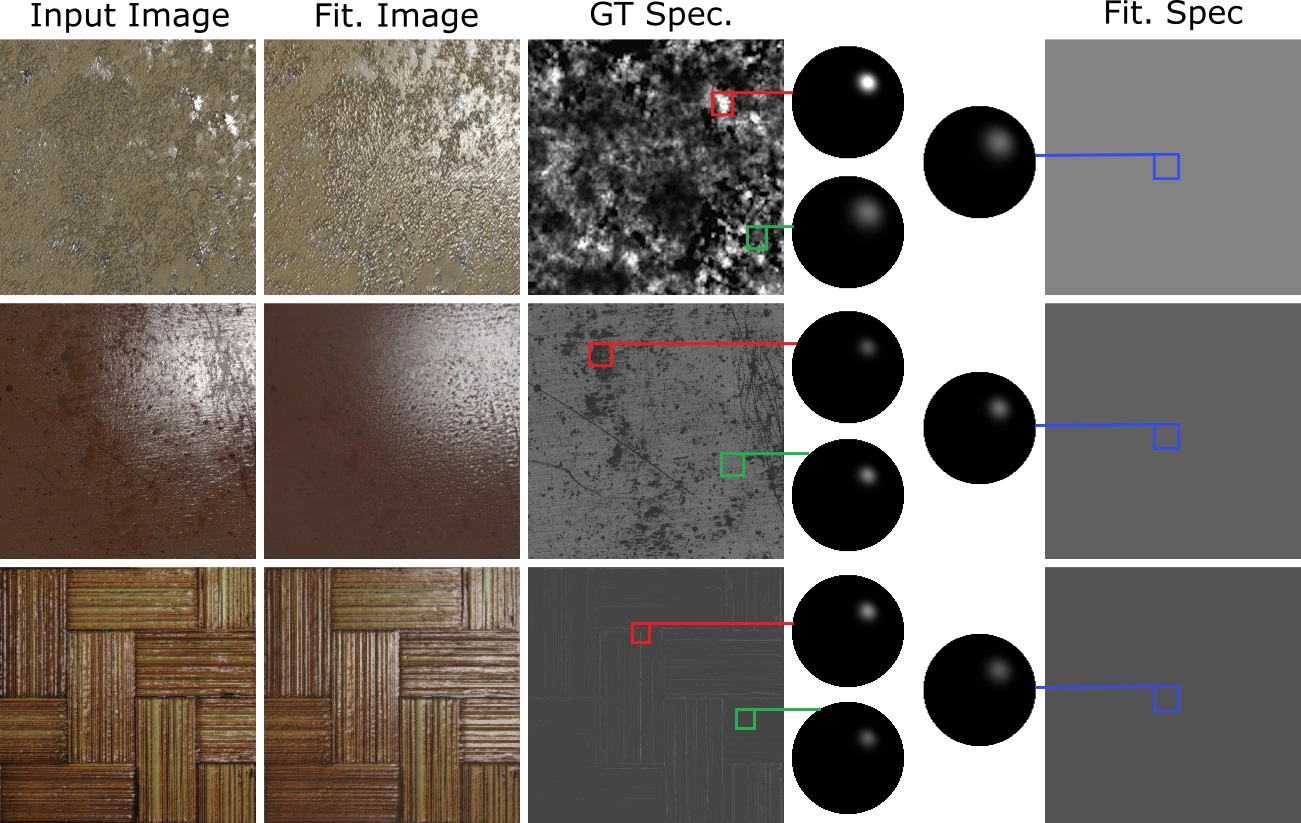}
  \vspace{-0.2cm}
  \caption{\textbf{Spatially Varying Specular Component} While
    SVBRDF-net expects and estimates a homogeneous specular
    component, it can still robustly estimate a plausible specular
    component when the spatial variations are modest.}
  \vspace{-0.2cm}
  \label{fig:spatialspec}
\end{figure}

Our SA-SVBRDF-nets are trained for a specific material type, and there
is no guarantee on the quality of the results when a input photograph
of a different material is provided. \autoref{fig:cross} shows a
result of feeding a spatially varying \emph{wood} material in the
\emph{metal} SA-SVBRDF-net, and vice versa. As expected, the resulting
cross estimation fails to produce physically plausible
results. Similarly, SA-SVBRDF-net expects spatially varying materials
with a homogeneous specular component. We empirically observe
(\autoref{fig:spatialspec}) that if the spatial variations in the
specular component are modest, SA-SVBRDF-net still produces a
plausible result. However, when faced with significant spatial
variations, the result is often unpredictable.

\section{Discussion}
\label{sec:disucssion}

Our results show that the proposed SA-SVBRDF-net can decompose a
photograph of a planar spatially varying sample under unknown distant
lighting in its reflectance components. Key to our method is the
self-augmentation training strategy. In this section, we further
explore the benefits, assumptions, and limitations of the proposed
self-augmentation procedure.

\paragraph{BRDF-net}
To validate the proposed self-augmentation training strategy we
introduce a CNN-based solution to the related, but less complex,
problem of estimating the homogeneous BRDF from a single image of a
smooth \emph{spherical} object under unknown natural lighting
conditions.  For validation purposes, we limit the experiment to
estimating monochrome BRDFs from a single monochrome image under
monochrome lighting.  The advantage of validating the
self-augmentation training strategy on a homogeneous BRDF-net is that
we can easily enumerate the full search space, allowing us to
synthesize labeled training data and visualize the learned and ground
truth manifolds and corresponding errors.

We employ the same network structure as the (log) roughness and
(log-relative) specular albedo prediction network in the previously
proposed SVBRDF-net. We express the relative specular albedo as
$\ks_{rel} = \ks / \kd$ instead of with respect to the average diffuse
albedo as in SVBRDF-net.

We generate a training set by uniformly sampling $10$ diffuse and
specular albedos in the $[0.05, 1.0]$ range, and $15$ roughness
samples log-uniformly sampled in $[0.02, 1.0]$, yielding a total of
$1500$ training samples. We also define a test set for evaluating the
errors by selecting samples mid-distance between any two consecutive
training samples; these are the sample points furthest away from the
training samples, which are expected to contain the largest error.

Because the specular-diffuse ratio estimated by the BRDF-net cannot be
used directly for synthesizing a provisional training image during
self-augmentation, we randomly select a diffuse/specular albedo pair
for rendering such that the posterior distribution is uniform.

\begin{figure*}[t]
  \hspace{-.7cm}
  \begin{minipage}{0.65\textwidth}
    \centering
    \includegraphics[width=1.0\textwidth]{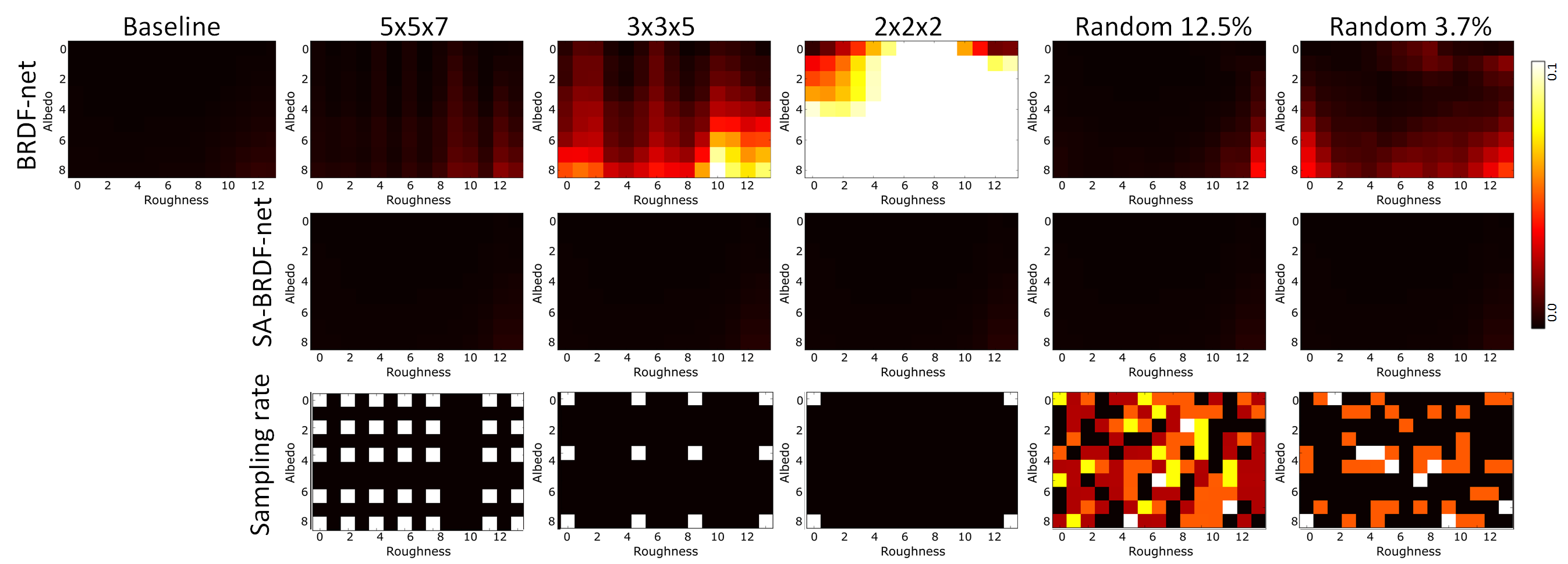}
  \end{minipage}
  \begin{minipage}{0.30\textwidth}
    \centering
    \small
    \begin{tabular}{|c|c|c|}
      \multicolumn{3}{c}{Total Mean Squared Error} \\
      \hline
       Subsample Rate   & BRDF-net & SA-BRDF-net   \\
      \hline
      Baseline          & $0.0004$ &               \\
      $5\times5\times7$ & $0.0010$ & $0.0004$      \\
      $3\times3\times5$ & $0.0060$ & $0.0005$      \\
      $2\times2\times2$ & $0.1001$ & $0.0005$      \\
      Random $12.5\%$   & $0.0007$ & $0.0004$      \\
      Random $3.7\%$    & $0.0029$ & $0.0004$      \\
      \hline
    \end{tabular}
  \end{minipage}
\caption{\textbf{Self-augmentation Validation on BRDF-net.} Error
  plots over the search space of (homogeneous) BRDF-net for different
  subsampling rates for the labeled training data. The reconstruction
  error is assessed on visualizations of the reference and estimated
  reflectance parameters. For visualization purposes, we integrate the
  3D search space over the specular albedo axis, and plot the
  resulting value for each diffuse albedo / specular roughness
  combination.  Top-left: baseline error on BRDF-net regressed over
  densely sampled training data. Top-row: the reconstruction error for
  different subsampling rates (uniform subsampling of diffuse albedo,
  specular albedo, and specular roughness at $5\times5\times7$,
  $3\times3\times5$ and $2\times2\times2$ sample rates, and randomly
  selecting samples that cover $12.5\%$ and $3.7\%$ of the search
  space). Middle-row: corresponding reconstruction error for
  SA-BRDF-net (i.e., trained with self-augmentation). Bottom-row:
  visualization of the sampling pattern. Right: total error over the
  full search space for each subsampling case.}
\label{fig:validation}
\end{figure*}

\paragraph{Validation}
To quantify the impact of self-augmentation on the accuracy of
BRDF-net, we regress a reference baseline BRDF-net on the full labeled
training set that densely and uniformly covers the full search space.
We measure the reconstruction error as the mean squared error between
visualizations of the ground truth and estimated reflectance
parameters under the same view and lighting conditions as the input
image.  Reducing the size of the training dataset when regressing the
BRDF-net without self-augmentation, dramatically decreases the
accuracy (\autoref{fig:validation} top row). We consider two different
subsampling strategies: uniformly reducing the sampling rate of
(diffuse and specular) albedo and roughness (i.e., effective sampling
rates: $5\times5\times7$, $3\times3\times5$, and $2\times2\times2$),
and randomly selecting a subset from the full training set (of size:
$12.5\%$, and $3.7\%$; see~\autoref{fig:validation} bottom row).
Compared to the baseline we observe a correlation in the error
distribution with the subsampling scheme -- regions with low training
data density exhibit larger error.  When using the same subsampled
training sets in conjunction with self-augmentation, where the
unlabeled image set equals the images left out from the labeled
training set, we observe an error rate similar to the baseline
BRDF-net (\autoref{fig:validation} middle row).  Surprisingly, even in
the extreme case where we only provide the corners of the parameter
space (i.e., $2\times2\times2$ sampling rate), the resulting
SA-BRDF-net still achieves accurate BRDF estimates.  

\begin{table*}[t]
%
%
\def\NA{\cellcolor{gray!10}}
%
\def\isoA{0.002000}  \definecolor{colA}{rgb}{0.75,0.00,0.00}
\def\isoB{0.001000}  \definecolor{colB}{rgb}{1.00,0.00,0.00}
\def\isoC{0.000700}  \definecolor{colC}{rgb}{1.00,0.50,0.10}
\def\isoD{0.000550}  \definecolor{colD}{rgb}{1.00,0.80,0.30}
\def\isoE{0.000400}  \definecolor{colE}{rgb}{1.00,1.00,0.70}
\newcommand{\CC}[1]{%
    \ifthenelse{\lengthtest{#1 pt > \isoA pt}}{ \cellcolor{colA}}{%
    \ifthenelse{\lengthtest{#1 pt > \isoB pt}}{ \cellcolor{colB}}{%
    \ifthenelse{\lengthtest{#1 pt > \isoC pt}}{ \cellcolor{colC}}{%
    \ifthenelse{\lengthtest{#1 pt > \isoD pt}}{ \cellcolor{colD}}{%
    \ifthenelse{\lengthtest{#1 pt > \isoE pt}}{ \cellcolor{colE}}{%
    }}}}}
    #1%
}
%
%
\caption{\textbf{Labeled-Unlabeled Training Data Ratio.} The error for different ratios of labeled and unlabeled data randomly sampled from the densely sampled search space.}\vspace{-0.2cm}
  \small
  \begin{center}
  \setlength{\fboxsep}{0.6mm} 
  \begin{tabular}{|c|c||c|c|c|c|c|c|c|c|c|}
    \hline
     Percent.& No Self-     & \multicolumn{9}{c|}{Percentage Unlabeled} \\
    \cline{3-11}
     Labeled & augmentation & 5        & 10           & 20            & 30            & 50            & 70            & 80            & 90            & 95            \\
    \hline
    5       & \CC{0.002549} & \CC{0.001395} & \CC{0.001141} & \CC{0.000884} & \CC{0.000689} & \CC{0.000704} & \CC{0.000651} & \CC{0.000578} & \CC{0.000592} & \CC{0.000628} \\
    10      & \CC{0.001252} & \CC{0.001382} & \CC{0.001027} & \CC{0.000720} & \CC{0.000760} & \CC{0.000671} & \CC{0.000584} & \CC{0.000634} & \CC{0.000592} & \NA           \\
    20      & \CC{0.000746} & \CC{0.001155} & \CC{0.000845} & \CC{0.000751} & \CC{0.000621} & \CC{0.000619} & \CC{0.000641} & \CC{0.000513} & \NA           & \NA           \\
    30      & \CC{0.000662} & \CC{0.000714} & \CC{0.000648} & \CC{0.000694} & \CC{0.000492} & \CC{0.000548} & \CC{0.000535} & \NA           & \NA           & \NA             \\
    50      & \CC{0.000562} & \CC{0.000660} & \CC{0.000559} & \CC{0.000552} & \CC{0.000506} & \CC{0.000470} & \NA           & \NA           & \NA           & \NA             \\
    70      & \CC{0.000619} & \CC{0.000601} & \CC{0.000462} & \CC{0.000550} & \CC{0.000499} & \NA           & \NA           & \NA           & \NA           & \NA             \\
    80      & \CC{0.000553} & \CC{0.000542} & \CC{0.000421} & \CC{0.000413} & \NA           & \NA           & \NA           & \NA           & \NA           & \NA             \\
    90      & \CC{0.000546} & \CC{0.000505} & \CC{0.000471} & \NA           & \NA           & \NA           & \NA           & \NA           & \NA           & \NA             \\
    95      & \CC{0.000550} & \CC{0.000471} & \NA           & \NA           & \NA           & \NA           & \NA           & \NA           & \NA           & \NA             \\
    100     & \CC{0.000499} & \NA          & \NA           & \NA           & \NA           & \NA           & \NA           & \NA           & \NA           & \NA             \\
    \hline
  \end{tabular}
 \end{center}
\label{tab:sampling}
\end{table*}

In the above experiment the sum of the labeled and unlabeled training
data densely covers the full search space. To better understand the
trade-offs between the number of labeled versus unlabeled training
data, we compare the error of different BRDF-nets trained on varying
ratios of labeled and unlabeled training data (uniformly) covering
different percentages of the densely sampled search space
(\autoref{tab:sampling}). Note, we consider $450$ randomly selected
lighting conditions per labeled training exemplar, and only a
\emph{single} randomly selected lighting condition per unlabeled
training exemplar -- increasing the number of lighting conditions for
each unlabeled training sample did not improve the error because
self-augmentation synthesizes a new provisional training exemplar
under a random lighting condition and thus automatically samples the
different lighting conditions already.  From this experiment we can
see that, as expected, the accuracy improves for increased number of
labeled and/or unlabeled training exemplars. However, care must be
taken when comparing the percentages of labeled versus unlabeled
training exemplars in~\autoref{tab:sampling} as these indicate
coverage of the search space and not number of images ($450$ lighting
conditions versus a single lighting condition for labeled and
unlabeled training exemplars respectively). Furthermore, we also
observe that self-augmentation is most advantageous when the number of
labeled training images is low; when the number of labeled training
images is large, the accuracy of SVBRDF-net is already good, and there
is only limited room for improvement. We argue that a significant
portion of the training data in regular CNN training merely aids in
refining the approximation of the search space, and only a small
portion is required to span the search space.  Consequently, care must
be taking to ensure that the selection of the labeled training data
exhibits sufficient diversity and fully spans the intended search
space, especially for very small labeled training sets.

\begin{figure*}[t]
\begin{minipage}{0.65\linewidth}
  \input{fig_smooth_space}
\end{minipage}
\hspace{0.3cm}
\begin{minipage}{0.32\linewidth}
  \input{fig_half}
\end{minipage}
\vspace{-0.1cm}
\end{figure*}

\paragraph{Practical Implications of Assumptions}
Self-augmentation assumes the search space is locally smooth, and no
``jumps'' or discontinuities occur in the space.  In other words, the
inverse of the targeted CNN should be well-defined. Intuitively,
unless the location of the discontinuity is precisely determined by
the labeled training data, the projection of the unlabeled data will
exhibit a large error depending to which side of the discontinuity the
estimate is biased toward. Depending on local gradients, it is
possible that self-augmentation drives the search space to an
(incorrect) local minimum. \autoref{fig:smooth_space} illustrates the
effect of a non-smooth search space on a simple 1D curve regression
with only two labeled training exemplars located the ends of the
range, and the unlabeled data distributed through the full space.  As
can be seen, while self-augmentation significantly improves the
accuracy, it tends to smooth out the discontinuity.

Self-augmentation also assumes that the unlabeled training data lies
in the region of the search space covered by the labeled training data
and no guarantees can be made on the accuracy outside the covered
search space (i.e., extrapolation).  However, empirically we found
that in many cases, self-augmentation also improves accuracy beyond
the covered region (\autoref{fig:extrapolation}). Consequently,
reliable estimates can still be obtained in practice as long as we
restrict unlabeled data and queries to lie close to the region covered
by the labeled data.

\paragraph{Limitations}
While the proposed SA-SVBRDF-net is able to recover plausible
reflectance parameters from a single photograph under unknown natural
lighting, the network is limited by the training data. Each
SA-SVBRDF-net is trained for a particular material
class. Consequently, for each new material type, a new network needs
to be trained.  Similarly, the quality of the appearance estimates
obtained with the proposed SA-SVBRDF-net are also limited by
information contained in the input photograph. If a particular
reflectance feature is not or limited present in the input image
(e.g., specular highlight), then the corresponding estimated
reflectance parameters will deviate more from the ground truth.

The proposed SA-SVBRDF-net is restricted to planar material samples
under distant lighting, and with a homogeneous specular component.
Generalizing SA-SVBRDF-net to unknown varying geometries and/or local
lighting conditions and/or spatially varying specularities is
non-trivial due to the ill-posed nature of the problem and the huge
search space.

While we have shown empirically that self-augmentation can greatly
reduce the required amount of labeled training data, a formal
theoretical derivation on the conditions for convergence, and the
conditions on the distribution of labeled/unlabeled data are missing.

\paragraph{Relation to other Deep Learning Methods}
The proposed method shares conceptual similarities to other recent
deep learning methods. Generative Adversarial Networks (GAN)
\cite{Goodfellow:2014:GAN} train two competing networks: a generating
network that synthesizes samples and a discriminative network that
attempts to distinguish real and synthesized samples.
Self-augmentation also relies on a ``synthesizer'', except that in our
case it is a fixed function. Similarly, Variational Auto-Encoders
(VAE)~\cite{Kingma:2013:AEV} train an end-to-end encoder and
decoder. The latent variables are parameterized with (Gaussian)
models, and variations of the search space are explored by sampling
these models.  The proposed solution also explores the search space by
random sampling of parameters (i.e., lighting and view). However, we
desire parameters with a precise physical meaning; the unsupervised
nature of VAEs make it difficult to generate models (i.e., parameters)
with physical meaning, even with conditional variants such as
CVAE~\cite{Sohn:2015:LSO} or CGAN~\cite{Liu:2016:CGA}.

\section{Conclusion}

We presented SA-SVBRDF-net, a convolutional neural network for
estimating physically plausible reflectance parameters from a single
photograph of a planar spatially varying material under unknown
natural lighting. Furthermore, we introduced a novel self-augmentation
training strategy to reduce the required amount of labeled training
data by leveraging the embedded information in a large collection of
\emph{unlabeled} photographs.  Our progressive self-augmentation
training strategy relies on the availability of the \emph{exact}
inverse of the desired SVBRDF-net in the form of a rendering
algorithm.  We demonstrated the effectiveness of our trained
SVBRDF-net, and thoroughly validated the self-augmentation training
strategy on a homogeneous BRDF-net.  For future work, we would like to
analyze the theoretical limits and conditions for self-augmentation,
as we believe that the proposed self-augmentation strategy is more
generally applicable beyond SVBRDF modeling. Furthermore, we would
like to investigate methods for generalizing the proposed SVBRDF-net
to non-planar material samples.

\section*{Acknowledgments}
We would like to thank the reviewers for their constructive feedback,
and the Beijing Film Academy for their help in creating the SVBRDF
datasets.  Pieter Peers was partially supported by NSF grant
IIS-1350323.

%
%
%
%
%
%

\vspace*{-0.5em}
\bibliographystyle{ACM-Reference-Format}
\bibliography{reference}

\end{document}